\documentclass[12pt, fleqn]{article}
\begin{document}

\title{\bf Orthogonality Effects in 
           Relativistic Models of Nucleon Knockout Reactions\thanks{
           This work is supported in part by the Natural Sciences
           and Engineering Research Council of Canada} }

\author{ {\bf J.I. Johansson and H.S. Sherif} \\
              Department of Physics, University of Alberta \\
              Edmonton,  Alberta, Canada T6G 2J1 \\
              and \\
              {\bf F. Ghoddoussi} \\
              Department of Physics, University of Alberta \\
              Edmonton,  Alberta, Canada T6G 2J1 \\
              and \\
              Department of Physics and Astronomy \\
              Wayne State University \\
              Detroit,  Michigan, USA 48202
       }

\date{\today}

\maketitle

\begin{abstract}

We study the effect of wave function orthogonality in the relativistic 
treatment of the nucleon removal reactions $\left(\gamma, p\right)$ and 
$\left(e, e^{\prime} p\right)$. 
The continuum wave function describing the outgoing nucleon is made orthogonal 
to the relevant bound states using the Gram-Schmidt procedure. 
This procedure has the advantage of preserving the asymptotic character of 
the continuum wave function and hence the elastic observables are unaffected.
The orthogonality effects are found to be negligible for 
$\left(e, e^{\prime} p\right)$ reactions
for missing momenta up to 700 $MeV/c$. 
This holds true for both parallel and perpendicular kinematics. 
By contrast the orthogonalization of the wave functions appears to have a more 
pronounced effect in the case of $\left(\gamma, p\right)$ reactions. 
We find that the orthogonality effect can be significant in this case 
particularly for large angles. 
Polarization of the outgoing protons and photon asymmetry show more sensitivity 
than the cross sections. 
If the orthogonality condition is imposed solely on this one hole state the 
effects are usually smaller.

\end{abstract}



\newpage

\section{Introduction}      \label{intro}

Nucleon removal reactions play a central role in clarifying important aspects 
of the nature of nuclear states, particularly those related to the 
shell structure of nuclei. 
There is currently renewed interest in these reactions when the probe is 
electromagnetic as more sophisticated electron beam machines are now available. 
The new generation of possible experiments opens the door to more precise and 
elaborate data that will help us understand the intricacies of the reaction 
mechanism and look at  possible changes in the structure of nucleons and other 
hadrons as they move within the nuclear environment. 
Spin dependent observables are expected to play a key role in these 
investigations.

Most earlier analyses of reactions such as the quasielastic 
$\left(e, e^{\prime} p\right)$  and $\left(\gamma, p\right)$ reactions 
have been carried out using the non-relativistic distorted wave 
impulse approximation (NRDWIA) \cite{IS94,LWW99}.  
In this approach the nucleon wave functions needed for the calculation of the 
reaction amplitude are solutions of the Schr\"{o}dinger equation,
and the current operators are obtained from the free probe-nucleon 
relativistic amplitude via some non-relativistic reduction scheme
\cite{MH62}.
Recently however it has become clear that a fully relativistic approach to the 
description of these reactions is more appropriate 
\cite{HJS95,JS96,Ud93,Ud95,JO94,LS88,JSL96}. 
This is evidenced by the improvement in the values of the spectroscopic 
factors obtained using this approach and indications of improvements 
to recoil polarization \cite{JS97,JS99} as well as cross sections at large 
missing momenta \cite{Ud96}. 

Because of the expectation that more precise and expanded data will become 
available shortly, there has been concentrated activity aimed at 
discerning details of the theoretical calculations of the amplitudes. 
The interplay between relativistic and non-relativistic approaches has been 
studied by several authors, and recently aspects of gauge dependence as well
as coupling effects have been investigated \cite{JK97,JK99}. 

The purpose of the present paper is to address one 
detail of the calculations of the amplitudes which is usually overlooked: 
namely, both relativistic and non-relativistic DWIA versions of the amplitude 
for a variety of reactions, including $\left(e, e^{\prime} p\right)$
and $\left(\gamma, p\right)$ 
(which are the focus of the present study) suffer from an orthogonality defect.
These distorted wave models utilize wave functions for bound and continuum 
states, which are solutions of equations containing different hamiltonians;
the bound states are generated from real potentials, while the continuum
states are generated from complex energy dependent optical potentials. 
The character of the latter potentials is dictated by the need to account for 
nucleon elastic scattering observables.
The wave functions are therefore not orthogonal, in contradiction to what a 
fundamental theory would require.

The impact of the nonorthogonality defect has been discussed in the past
and ways to remedy it have been suggested 
\cite{AW77,No78,ENW79,CS79,AGS83,Bo82,Bo83,FHK72}. 
The majority of these discussions were done in a non-relativistic framework 
but many of the remedies suggested have never been utilized in 
main stream calculations.
Noble \cite{No78} attempted to recast the amplitude for photoabsorption 
into a form that takes into account the orthogonality of the nucleon 
wave functions. 
Difficulties with this approach have been discussed by 
Eisenberg, Noble and Weber \cite{ENW79}.
Celenza and Shakin \cite{CS79} suggested that a modified hamiltonian 
for the continuum states can be constructed in such a manner as to 
guarantee both orthogonality with the bound states, and a fit to 
the proton elastic scattering data. 
This proposition was never tested in practice. 
Other possibilities discussed by Boffi {\it et al.} \cite{Bo82,Bo83} 
involved the introduction of effective operators, but it turned out 
that these were too difficult to calculate. 

For the two reactions under consideration, however, a more practical approach 
was identified and used to assess the role of non-orthogonality on the 
non-relativistic amplitudes.  
It was suggested by Boffi {\it et al.} \cite{Bo82,Bo83} as well as by 
Ciofi Degli Atti {\it et al.} \cite{AGS83}  that the continuum states can be made 
orthogonal to the bound states by applying a 
Gram-Schmidt orthogonalization procedure.
In the present work we use this same method to examine the importance of 
orthogonalization in the relativistic approach to nucleon knockout reactions.

In the following text we give a brief outline of the relativistic calculations
for nucleon knockout induced by the quasi elastic scattering of electrons or 
the absorption of real photons. 
We then introduce the method of orthogonalization of the nucleon wave 
functions and calculate the resulting corrections to the reaction amplitude. 
A discussion of the impact of these changes on the reaction observables is 
given in section 3. 
We conclude in section 4.

\section{The Relativistic Calculations}   \label{rel_obs}

The relativistic calculations of the amplitude, in the one photon
exchange model for the $\left(e, e^{\prime}p\right)$ process,
are discussed in Refs. \cite{HJS95,JS96,Ud93,Ud95,JO94}.
Relativistic calculations for $\left(\gamma, p\right)$ reactions 
are discussed in Refs. \cite{JS96,LS88,JSL96}.
It is sufficient for the purpose of the present discussion to focus on 
the nuclear matrix element portion of the amplitude, which can be written 
in a unified form for reactions involving either virtual or real photons. 
In the notation of Ref. \cite{JS96} the relativistic expression for the 
nuclear matrix element $N_{\beta}^{\mu M_B}$,  
leading to a specific final state of the residual nucleus 
for both the $\left(\gamma, p\right)$ and
$\left(e, e^{\prime} p\right)$ reactions can be written as
\begin{eqnarray}
  N_{\beta}^{\mu M_B} = \int d^3r \; 
        \left[ \Psi_{\mu}^{\left( - \right)} \right]^\dagger 
           \left( \mbox{\boldmath{$p$}}, \mbox{\boldmath{$r$}} \right)
        \Gamma_{\beta}
        \Psi_{J_{B}, M_{B}} \left( \mbox{\boldmath{$r$}} \right)
        \exp \left( i \mbox{\boldmath{$q$}} \cdot
                      \mbox{\boldmath{$r$}} \right)     ,
   \label{nuc_mat}
\end{eqnarray}
where $M_B$ and $\mu$ are the spin projections of the bound and continuum 
nucleons, respectively, and $\mbox{\boldmath{$p$}}$ is the 3-momentum 
of the latter. 
The $4 \times 4$ matrix $\Gamma_{\beta}$, operating on the nucleon spinors,
is given in detail for the $\left(e, e^{\prime} p\right)$ reaction, 
in Eq. (2.8) of reference \cite{HJS95}.
For the $\left(\gamma, p\right)$ reaction, $\Gamma_{\beta}$ is written
explicitly in Ref. \cite{JS97_i}.
 
In relativistic DWIA (RDWIA) calculations of the amplitude, 
$\Psi_{\mu}$ is chosen as the Dirac continuum wave function in the 
elastic channel. 
It is generated by using a complex nucleon-nucleus optical potential, 
and is required to give an acceptable account of the elastic data. 
The spinor $\Psi_{J_{B}, M_{B}}$, on the other hand, is a solution of the 
bound state Dirac equation, which contains real scalar and vector potentials. 
It is clear then that the nucleon wave functions used in the RDWIA are 
not orthogonal since they are solutions of the Dirac equation for two 
different hamiltonians. 
It is the lack of orthogonality of the spinors  $\Psi_{\mu}$ and 
$\Psi_{J_{B}, M_{B}}$ that is the subject of the present study.

\subsection{Orthogonal States}   \label{orthog_def}

The orthogonal continuum wave function is obtained via the Gram-Schmidt 
method.
This involves subtracting a multiple of the bound state wave functions for 
each of the occupied levels of the target nucleus.
The approach has the advantage of being easily incorporated within the 
RDWIA approach as a means of gaining some insight into the possible corrections 
arising from the orthogonality defect. 
The proposed orthogonal wave function can be written in terms of the original 
distorted wave and the bound state wave functions as
\begin{eqnarray}
  \tilde{\Psi}_{\mu}^{\left( + \right)} \left( \mbox{\boldmath{$p$}}, 
                                               \mbox{\boldmath{$r$}} \right)
     = \Psi_{\mu}^{\left( + \right)} \left( \mbox{\boldmath{$p$}}, 
                                            \mbox{\boldmath{$r$}} \right)
       - \sum_{{L_{B}^{\prime}, J_{B}^{\prime}, M_{B}^{\prime}}} 
            {\beta^{{L_{B}^{\prime}, J_{B}^{\prime}, M_{B}^{\prime}}}_{\mu}
             \Psi_{{L_{B}^{\prime}, J_{B}^{\prime}, M_{B}^{\prime}}} 
                  \left( \mbox{\boldmath{$r$}} \right)
            }      ,
   \label{orthog_cont}
\end{eqnarray}
where the sum is over all the occupied bound states (characterized here by their
angular momentum labels $L_{B}^{\prime}, J_{B}^{\prime}, M_{B}^{\prime}$).
Note that the coefficient, 
$\beta^{L_{B}^{\prime}, J_{B}^{\prime}, M_{B}^{\prime}}_{\mu}$, 
in the second term of Eq. (\ref{orthog_cont}) is 
required to have the same asymptotic spin projection $\mu$, as the original 
continuum wave function.  

It was shown by Ciofi Degli Atti {\it et al.} \cite{AGS83} that there is 
a close connection between the above form of orthogonalized wave 
functions and  the requirement of 
antisymmetrization of the final state wave function in reactions initiated 
by the real photons. 
A crucial advantage of the above prescription is that the resulting orthogonal 
continuum wave functions continue to yield the same proper description of 
elastic scattering provided by the original proton continuum wave function.

The expansion coefficients 
$\beta^{{L_{B}^{\prime}, J_{B}^{\prime}, M_{B}^{\prime}}}_{\mu}$ 
are determined by requiring that the wave function given in 
Eq. (\ref{orthog_cont}) is orthogonal to all occupied states of the target; 
this yields
\begin{eqnarray}
  \beta^{L_{B}, J_{B}, M_{B}}_{\mu}
     = \left< \Psi_{L_{B}, J_{B}, M_{B}} \left( \mbox{\boldmath{$r$}} \right) 
              {\mbox{\large{$|$}}}
              \Psi_{\mu} \left( \mbox{\boldmath{$p$}}, 
                                \mbox{\boldmath{$r$}} \right)
            \right>      .
   \label{exp_coeff_1}
\end{eqnarray}

The detailed evaluation of these expansion coefficients requires the standard
partial wave expansion for the wave functions.
For the continuum state we write
\begin{eqnarray}
   \psi^{\left( + \right)}_{\mu} 
       \left( \mbox{\boldmath{$p$}}, \mbox{\boldmath{$r$}} \right)
     &=& 4 \pi \left[ \frac{E_{N} + m_{N}}{2 m_{N}} \right]^{1/2}
      \nonumber \\
        & & 
           \sum_{L J M}{i^{L}}
           \left[ {Y_{L}^{M - \mu}}
           \left( \widehat{\mbox{\boldmath{$p$}}} \right) \right]^{\ast}
           \left( L, 1/2; M - \mu, \mu| J, M \right)
      \nonumber \\
        & & \hspace{8 mm} \times
           \left[ \matrix{  f_{L J} \left( r \right)        \cr
                            - i g_{L J} \left( r \right) 
                            \mbox{\boldmath{$\sigma$}} \cdot
                            \widehat{\mbox{\boldmath{$r$}}}     \cr
                         } \right]
           {\cal Y}_{L \; \; 1/2 \; \; J}^{\; \; \; M}
                 \left( \widehat{\mbox{\boldmath{$r$}}} \right)  .
   \label{cont_wf}
\end{eqnarray}
where the upper and lower component radial functions,
$f_{L J} \left( r \right)$ and $g_{L J} \left( r \right)$,
are solutions of equations containing complex scalar and vector potentials.
For the bound state with angular momentum quantum numbers
$L_{B}, J_{B}, M_{B}$ we can write
\begin{eqnarray}
\Psi_{J_{B}, M_{B}} \left( \mbox{\boldmath{$r$}} \right)
   = \left[ \matrix{  f_{L_{B} J_{B}} \left( r \right)        \cr
                            - i g_{L_{B} J_{B}} \left( r \right) 
                            \mbox{\boldmath{$\sigma$}} \cdot
                            \widehat{\mbox{\boldmath{$r$}}}     \cr
                         } \right] 
     {\cal Y}_{L_{B} \; 1/2 \; J_{B}}^{\; \; \; M_{B}}
     \left( \widehat{\mbox{\boldmath{$r$}}} \right)   .
   \label{bound_wf}
\end{eqnarray}
Using Eqs. (\ref{cont_wf}) and (\ref{bound_wf}) in Eq. (\ref{exp_coeff_1}), 
we get for the expansion coefficients
\begin{eqnarray}
  \beta^{L_{B}, J_{B}, M_{B}}_{\mu}
     = & & 4 \pi \left[ \frac{E_{N} + m_{N}}{2 m_{N}} \right]^{1/2}
           \left( L_{B}, 1/2; M_{B} - \mu, \mu| J_{B}, M_{B} \right)
      \nonumber \\      
        & & 
           \left[ {Y_{L_{B}}^{M_{B} - \mu}}
           \left( \widehat{\mbox{\boldmath{$p$}}} \right) \right]^{\ast}
           b_{L_{B}, J_{B}}     .
   \label{exp_coeff_2}
\end{eqnarray}
where $b_{L_{B}, J_{B}}$ is the overlap integral between the bound
and distorted wave radial functions:
\begin{eqnarray}
  b_{L_{B}, J_{B}}
     = \int{r^{2} dr} 
               \left[   f_{L_{B} J_{B}}^{(bound)} \left( r \right)
                        f_{L_{B} J_{B}}^{(cont)} \left( r \right)
                      + g_{L_{B} J_{B}}^{(bound)} \left( r \right)
                        g_{L_{B} J_{B}}^{(cont)} \left( r \right)
                    \right]     .
   \label{overlap_coeff}
\end{eqnarray}

Using Eqs. (\ref{cont_wf}) and (\ref{exp_coeff_2}) in 
Eq. (\ref{orthog_cont}) allows us to write a partial wave expansion for the
distorted wave which will be orthogonal to the bound states.
The result is that the continuum wave function is changed only in 
partial waves with angular momentum quantum numbers corresponding to
the occupied levels of the target nucleus.
The resulting continuum wave function is 
suitable for use in calculations of the nuclear matrix
element of Eq. (\ref{nuc_mat}) without further modification,
and is straightforward to include in existing reaction codes.

\section{Discussion}         \label{disc}  

To assess the impact of the wave function nonorthogonality in realistic situations 
we have carried out calculations for a number of cases involving photonuclear knockout 
of protons for photon energies above 60 $MeV$. 
We also extend the study to electron quasielastic scattering on the same nuclei.
In these calculations we use bound state wave functions which are generated using 
phenomenological Dirac scalar and vector potentials with Woods-Saxon shapes. 
The parameters of the potentials are determined by the requirement that the 
separation energies of the states for which we have experimental data are
reasonably well reproduced. 
This binding potential is used to generate the entire family of occupied 
states in the target.
The distorted waves are generated using the energy dependent potentials of 
Cooper {\it et al.} \cite{COPE}. 
For the $^{16}O$ target we use the energy--dependent potentials specific to 
this target (EDAI O16), while for the other nuclei considered we use the 
energy-- and mass--dependent set (EDAD Fit 1)
These distorted wave functions are then made orthogonal to the family of 
bound states following the procedure discussed in section 2.

The size of the orthogonality effect is linked to the magnitude of the 
coefficients $b_{L_{B}, J_{B}}$ given in Eq. (\ref{overlap_coeff}). 
As is seen from this equation, the coefficients are radial overlaps 
between bound and continuum wave functions. 
Figure 1 shows the behavior of the overlap coefficient 
for the $1p_{1/2}$ state in 
$^{16}O$ as a function of the energy of the continuum nucleon. 
Part (a) of the figure shows the real (solid) and imaginary (dashed)
parts of the coefficient on a standard plot, while part (b) shows the 
absolute value of the real and imaginary parts on a semilogarithmic plot.
As one might expect these overlap coefficients are larger at lower energies 
and drop quickly as the energy increases. 
They have oscillatory behavior reflecting that particular character of the 
continuum radial wave functions. 
At low energies ($\approx$10 $MeV$) the magnitude 
of the coefficients can be close to 0.1.  
Although these are relatively large, they are not likely  to  have large 
effects in the energy region of interest for the reactions under consideration. 
Note also that the orthogonality requirement affects only a limited number of 
partial waves (depending on the target nucleus) in the continuum wave function:
those corresponding to occupied states in the target nucleus.

In the present work, the effects of orthogonality are studied in two cases: 
\begin{itemize}
\item[i)]
  {\em Minimal orthogonality}: the continuum wave function is made orthogonal 
  to only the one-hole state involved in the reaction; 
\item[ii)]
  {\em Full orthogonality}: the continuum wave function is made orthogonal 
  to all occupied states of the target. 
\end{itemize}
In the following we give an account of the results obtained for the two 
reactions under consideration.

\subsection{$\left(\gamma, p\right)$ Reactions:}
 
We have investigated the effect of non-orthogonality on the cross section, 
proton polarization and photon asymmetry in 
$\left(\gamma, p\right)$ reactions on four nuclei: $^{12}C$, $^{16}O$
$^{40}Ca$ and $^{208}Pb$, for photon energies from 60 to 500 $MeV$. 
Two representative examples of these calculations are shown in Figs. 2 and 3. 
In Fig. 2 we show the calculations for the reaction 
$^{16}O\left(\gamma, p\right)$ at an incident photon energy of 
$E_{\gamma} = 100$ $MeV$.
A proton is removed from the $1p_{1/2}$ level and the residual nucleus is
left in the ground state.
In Fig. 3 we show similar calculations for incident photon energy 
$E_{\gamma} = 196$ $MeV$.
To put the comparisons in a proper context we also show the experimental
cross section data for the above reactions \cite{FO77,Ad88}.

We note that the effects of minimal orthogonalization are generally smaller 
than those for full orthogonalization. 
The cross sections (part (a) of the figures) are affected only slightly 
at most angles but the deviations can be large at extreme backward angles. 
The relative increase of the effect at large angles is a reflection of the fact, 
mentioned above, that the orthogonality requirement modifies only low partial waves. 
The effects on proton polarization (part (b)) and photon asymmetry (part (c)) 
are noticeable over most of the angular range, 
but are small at the very forward angles. 
This is an indication that these two observables are somewhat more sensitive 
to the orthogonality defect than are the cross sections.

Figures 4 and 5 shed more light on the role of orthogonality and the 
dependence on the photon energy and the angle of the outgoing protons. 
Here we show the photon energy dependence of the cross section on 
$^{40}Ca$ at proton angles of $\theta_{p} = 90^{\circ}$ (Fig. 4) 
and $\theta_{p} = 135^{\circ}$ (Fig. 5). 
The cross section data shown in both figures are from Leitch {\em et al.} 
\cite{Le86}. 
We observe from Fig. 4 that at  $\theta_{p} = 90^{\circ}$ the effects on the 
direct knockout cross section are rather small up to photon energies of 300 $MeV$ 
(calculations show this feature still prevails at higher energies). 
At $\theta_{p} = 135^{\circ}$ (Fig. 5), we find larger effects on the 
cross section, particularly when full orthogonalization is imposed on 
the wave functions. 
These results are in qualitative agreement with the observations made
by Ciofi Degli Atti {\it et al.} \cite{AGS83} using the non-relativistic approach.
Parts (b) and (c) of the figures show the effects on 
proton polarization and photon asymmetry, respectively. 
Here we note large effects on both observables at both angles and over 
the entire energy range 
(these effects are much smaller at $\theta_{p} = 45^{\circ}$). 
It is  evident that the spin observables are more sensitive to the 
orthogonality of the wave functions.
  
Boffi {\it et al.} \cite{Bo82,Bo83} argued that the effect of non-orthogonality 
on the unfactorized $\left(\gamma, p\right)$ cross sections vanishes because of 
the transverse nature of the electromagnetic current. 
Our present calculations do not support this conclusion; as seen in the preceding 
figures, we do observe some effects when using only minimal orthogonalization. 
The reason for the difference is that in Eq. (\ref{orthog_cont}), 
even if one restricts the sum to a single orbital ($L_{B}, J_{B}$) 
there are still contributions from different magnetic substates. 
As a result some of the contributions from orthogonalization do not vanish. 
This difference can be stated as due to the possibility of spin-flip in the 
outgoing channel (which is closely linked to the presence of the spin-orbit 
interaction); 
such a possibility was not taken into account by Boffi 
{\it et al.}.

Earlier estimates of the orthogonality effects in $\left(\gamma, p\right)$ 
reactions were obtained for non-relativistic calculations by 
Fink {\it et al.} \cite{FHK72} and for 
relativistic calculations by Lotz \cite{Lo89}. 
The procedure followed was to replace the distorting potential of 
the continuum proton with the bound state potential.
The continuum and bound state hamiltonians are thus identical and hence
the wave functions are orthogonal. 
The dotted curve in Fig. 4 shows the calculated observables when this procedure 
is implemented. 
The effects are seen to be much more exaggerated in this case. 
It should be noted that one important shortcoming of this procedure is 
that the new distorted wave function is no longer able to reproduce the 
elastic scattering data. 
This difficulty is also present in non-relativistic calculations that use 
continuum RPA wave functions \cite{SRW96}.

\subsection{$\left(e, e^{\prime} p\right)$ Reactions:}

We have also carried out similar calculations for the 
$\left(e, e^{\prime} p\right)$ reaction.
Boffi {\it et al.} \cite{Bo82,Bo83} and Ciofi Degli Atti {\it et al.} 
\cite{AGS83} have reported that in the 
non-relativistic calculations of the amplitudes the orthogonalization effects
are quite small. 
Our relativistic calculations confirm this conclusion. 
We have carried out calculations for electron energies of 460 $MeV$ and 2.4 
$GeV$ and found that the spectral functions are affected only slightly by 
the imposition of orthogonality. 
Our calculations covered missing momenta in the range $\pm$700 $MeV/c$. 
At proton energies of 100 $MeV$ or higher the effects of orthogonality are
essentially negligible.
At low proton kinetic energies we find some orthogonality effects
in the proton polarization calculations. 
Figure 6 shows calculations for the reaction
$^{16}O\left(e, e^{\prime} p\right)$ in perpendicular kinematics with
a proton energy of $T_{p} = 35$ $MeV$.
It is evident in this figure that orthogonality effects are 
small except at the large missing momenta. 
Figure 7 shows quasifree electron scattering calculations for the $^{40}Ca$
target in parallel kinematics, with a proton energy of $T_{p} = 100$ $MeV$.
The orthogonality corrections due to Eq. (\ref{orthog_cont}) are negligible,
but replacing the distorting potential with the binding potential, as was done
in Fig. 4, has a large effect.
This prescription cannot be advocated because the distorted wave is no longer
able to describe proton elastic scattering.
Note that these latter calculations cannot describe the electron 
scattering data very well either.

\section{Conclusion}

In this paper we have investigated the role of wave function orthogonality in 
relativistic calculations of $\left(\gamma, p\right)$ and 
$\left(e, e^{\prime} p\right)$ reactions. 
We have used a Gram-Schmidt procedure \cite{Bo82,Bo83,AGS83} to orthogonalize 
the continuum and bound-state wave functions of the nucleon. 
This procedure, besides being closely linked to the antisymmetrization 
\cite{AGS83} of the final state wave function, also has the merit of 
leaving the continuum wave function asymptotically intact. 
This ensures that the wave function continues to provide the same good description 
of the elastic scattering data as the original distorted wave. 
Calculations were carried out for two schemes, one where orthogonality is 
imposed only in relation to the specific hole state involved in the reaction, 
and full orthogonality with all occupied levels. 
The effects in the latter case were found to be larger.

The present study suggests that the orthogonality effects are not a 
serious problem for $\left(e, e^{\prime} p\right)$ reactions. 
Calculations over a wide range of missing momenta do not show large
deviations from the traditional nonorthogonal calculations except for
the spin dependent observables at large missing momenta.

The orthogonality effects are found to be non-negligible in the case
of $\left(\gamma, p\right)$ reactions. 
Our calculations show that these effects tend to be more pronounced 
at large angles of the outgoing protons.
The spin observables appear to be more sensitive to the orthogonality
of the wave functions. This may have important ramifications for
calculations of the amplitudes for higher order processes such as
those associated with two-nucleon currents.

\newpage

\section* {Figure Captions}

\noindent FIG. 1. The overlap coefficients, $b_{L_{B}, J_{B}}$, 
of Eq. (\ref{overlap_coeff}) for the $1s_{1/2}$ state in the $^{16}O$ target.
The binding potentials have Woods-Saxon shapes, 
while the proton optical potentials are the E-dependent optical potential 
for $^{16}O$ from Ref. \cite{COPE}.
(a) linear plot of real (solid) and imaginary (dashed) parts, and, 
(b) semilogarithmic plot of absolute value of real (solid) and imaginary (dashed) parts.

\vspace{4 mm}
\noindent FIG. 2. Effect of wave function orthogonalization in
the $^{16}O\left(\gamma, p\right)^{15}N_{g.s.}$ reaction
involving the knockout of a $1p_{1/2}$ proton.
The energy of the incident photon is 100 $MeV$.
The binding potentials have Woods-Saxon shapes, 
while the proton optical potentials are the E-dependent optical potentials 
for $^{16}O$ from Ref. \cite{COPE}.
(a) Cross section.
(b) Proton polarization.
(c) Photon asymmetry.
Solid curves  --- no orthogonality.
Long--dashed curves --- minimal orthogonality.
Short--dashed curves --- full orthogonality.
The data are from Ref. \cite{FO77}.

\vspace{4 mm}
\noindent FIG. 3. Same as Fig. 2, except the 
energy of the incident photon is 196 $MeV$.
The data are from Ref. \cite{Ad88}.

\vspace{4 mm}
\noindent FIG. 4. Effect of wave function orthogonalization in
the $^{40}Ca\left(\gamma, p\right)^{39}K_{g.s.}$ reaction
involving the knockout of a $1d_{3/2}$ proton.
The angle of the final proton is $90^{\circ}$.
The binding potentials have Woods-Saxon shapes, 
while the proton optical potentials are E- and A-dependent optical potentials 
(Fit 1) from Ref. \cite{COPE}.
(a) Cross section.
(b) Proton polarization.
(c) Photon asymmetry.
Solid curves  --- no orthogonality.
Long--dashed curves --- minimal orthogonality.
Short--dashed curves --- full orthogonality.
Dotted curves --- continuum wave function calculated using binding potential.
The data are from Ref. \cite{Le86}.

\vspace{4 mm}
\noindent FIG. 5. Same as Fig. 4, except the 
angle of the final proton is $135^{\circ}$.
The data are from Ref. \cite{Le86}.

\vspace{4 mm}
\noindent FIG. 6. Effect of wave function orthogonalization in
the $^{16}O\left(e, e^{\prime} p\right)^{15}N_{g.s.}$ reaction
involving the knockout of a $1p_{1/2}$ proton.
The energy of the incident electron is 456 $MeV$,
with perpendicular kinematics,
and the proton kinetic energy is $T_{p} = 35$ $MeV$.
The binding potentials have Woods-Saxon shapes, 
while the proton optical potentials are the E-dependent optical potentials 
for $^{16}O$ from Ref. \cite{COPE}.
(a) Cross section.
(b) Proton polarization.
Solid curves  --- no orthogonality.
Long--dashed curves --- minimal orthogonality.
Short--dashed curves --- full orthogonality.

\vspace{4 mm}
\noindent FIG. 7. Effect of wave function orthogonalization in
the $^{40}Ca\left(e, e^{\prime} p\right)^{39}K_{g.s.}$ reaction
involving the knockout of a $1d_{3/2}$ proton.
The energy of the incident electron is 460 $MeV$,
with parallel kinematics,
and the proton kinetic energy is $T_{p} = 100$ $MeV$.
The binding potentials have Woods-Saxon shapes, 
while the proton optical potentials are E- and A-dependent 
optical potentials (Fit 1) from Ref. \cite{COPE}.
(a) Cross section.
(b) Proton polarization.
Solid curves  --- no orthogonality.
Long--dashed curves --- minimal orthogonality.
Short--dashed curves --- full orthogonality.
Dotted curves --- continuum wave function calculated using binding potential.
The data are from Ref. \cite{Kr89}.

\begin{figure}
\begin{picture}(1100,400)(0,0)
\includegraphics{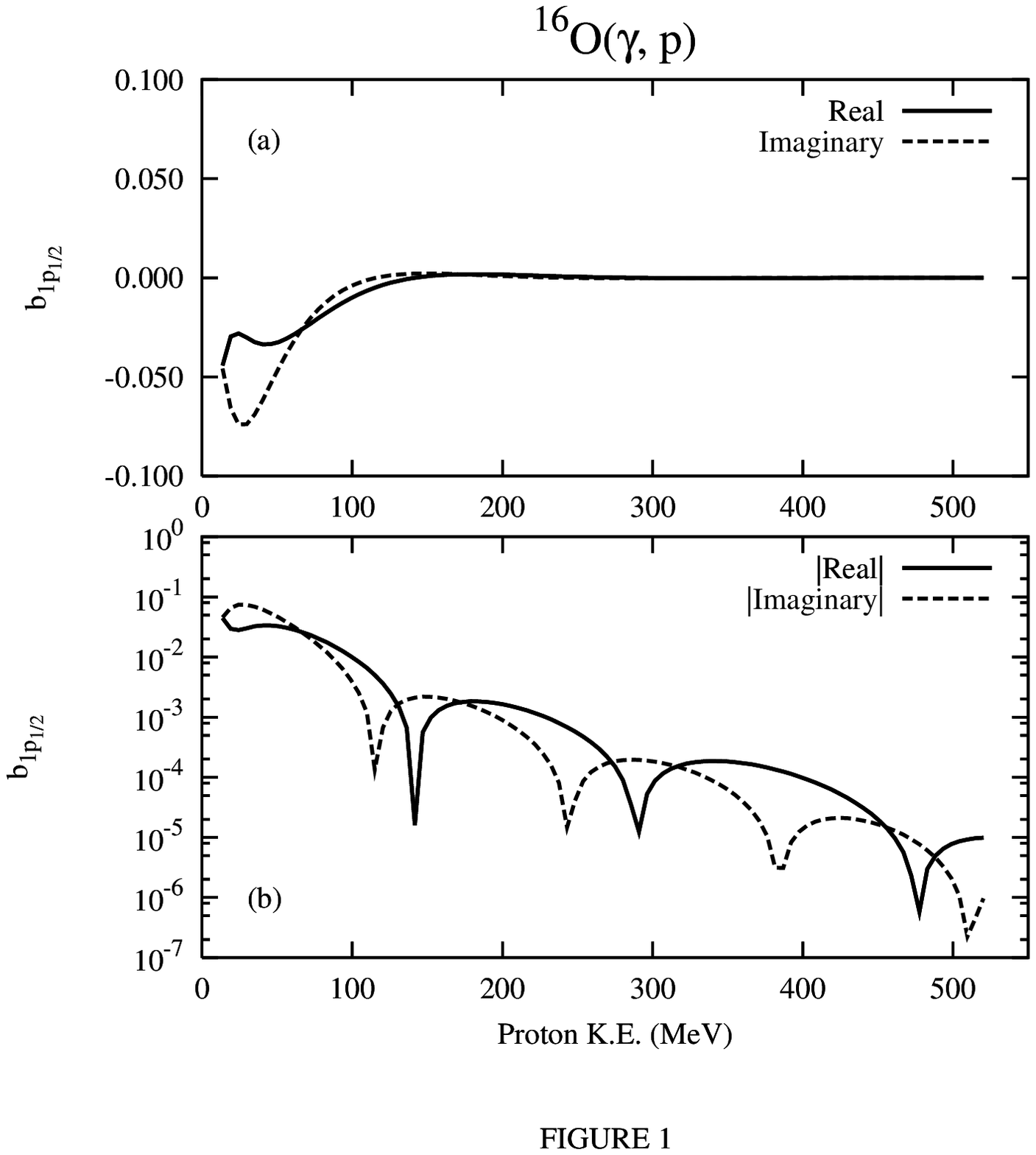}
\end{picture}
\end{figure}

\begin{figure}
\begin{picture}(1100,400)(0,0)
\includegraphics{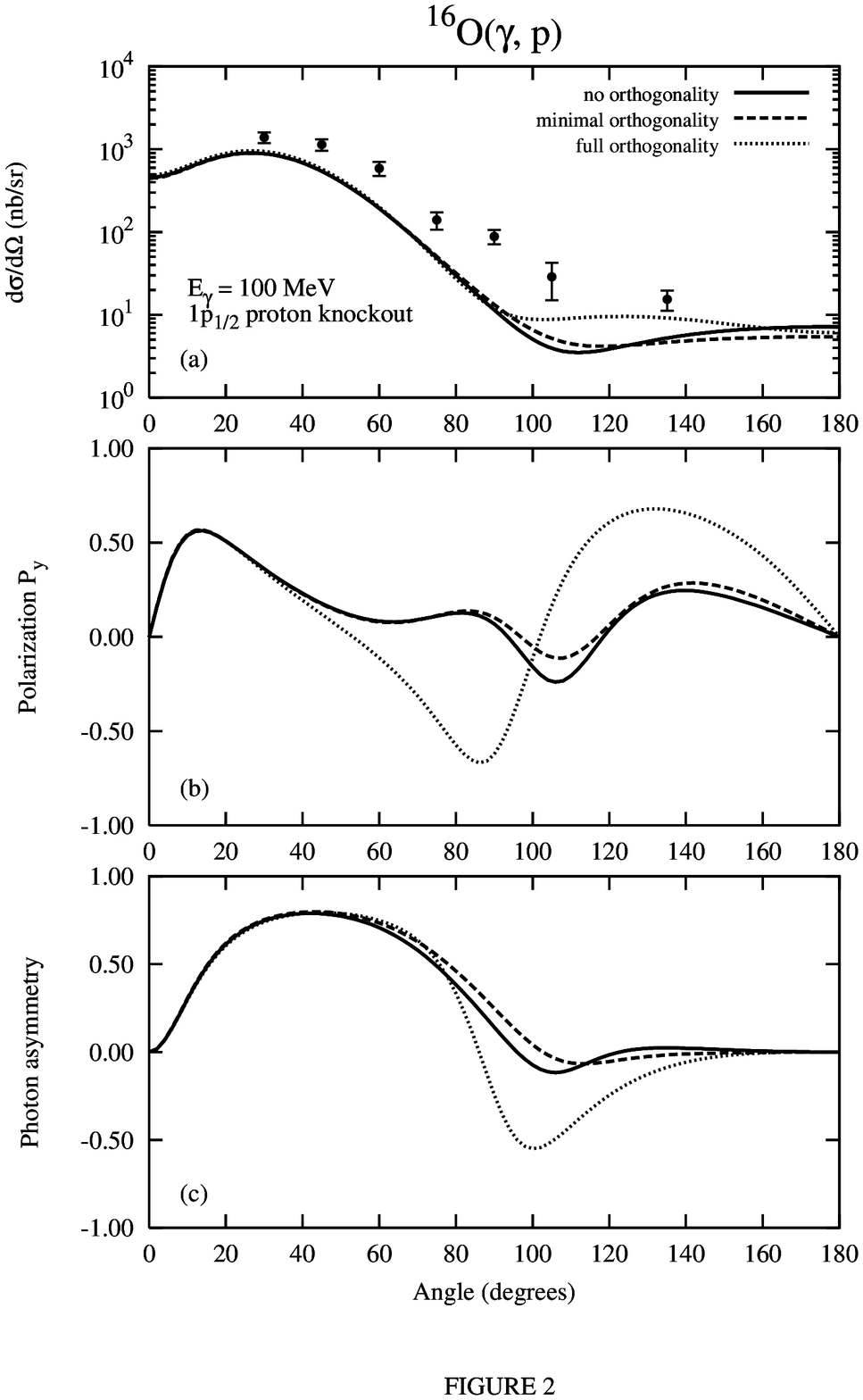}
\end{picture}
\end{figure}

\begin{figure}
\begin{picture}(1100,400)(0,0)
\includegraphics{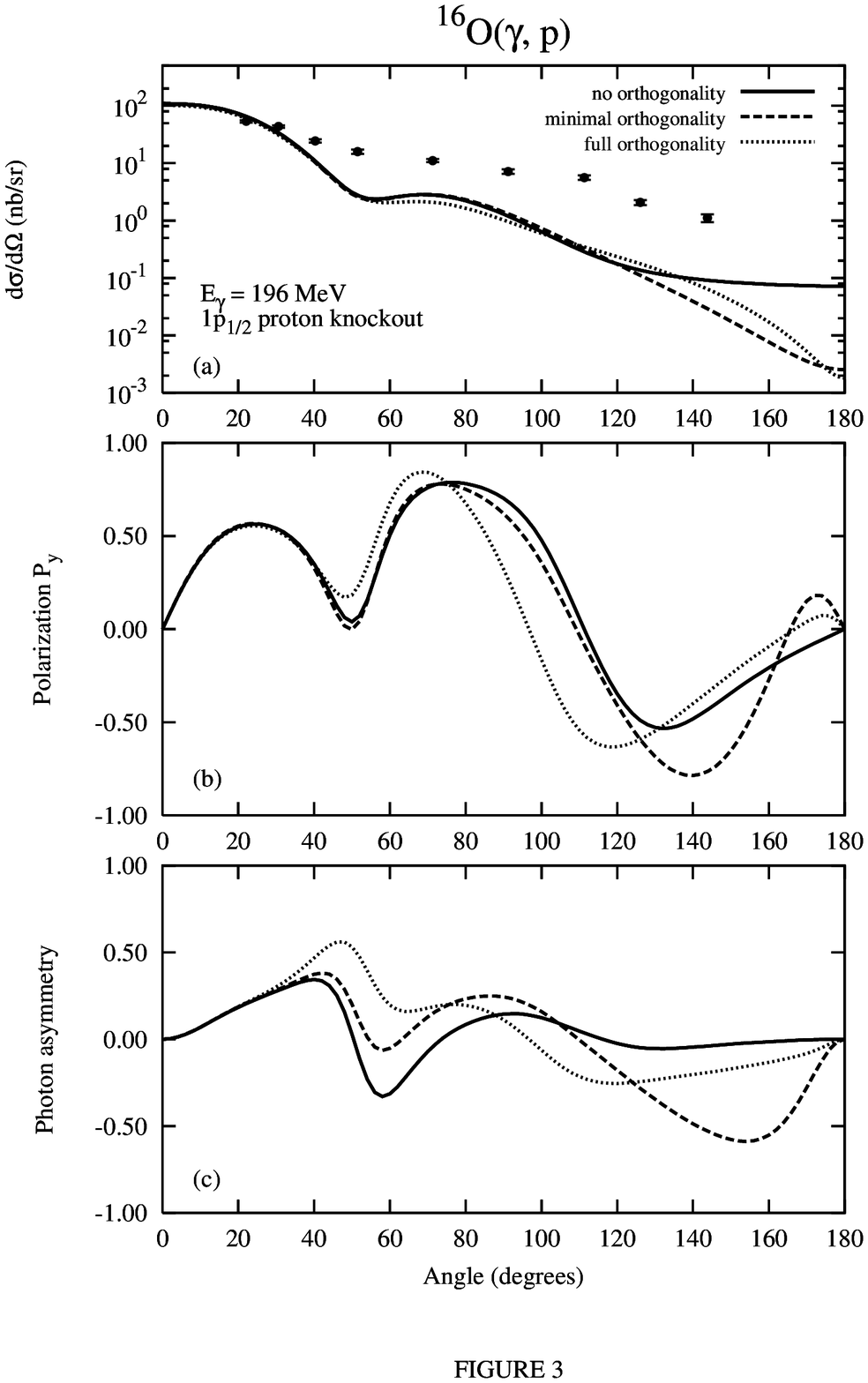}
\end{picture}
\end{figure}

\begin{figure}
\begin{picture}(1100,400)(0,0)
\includegraphics{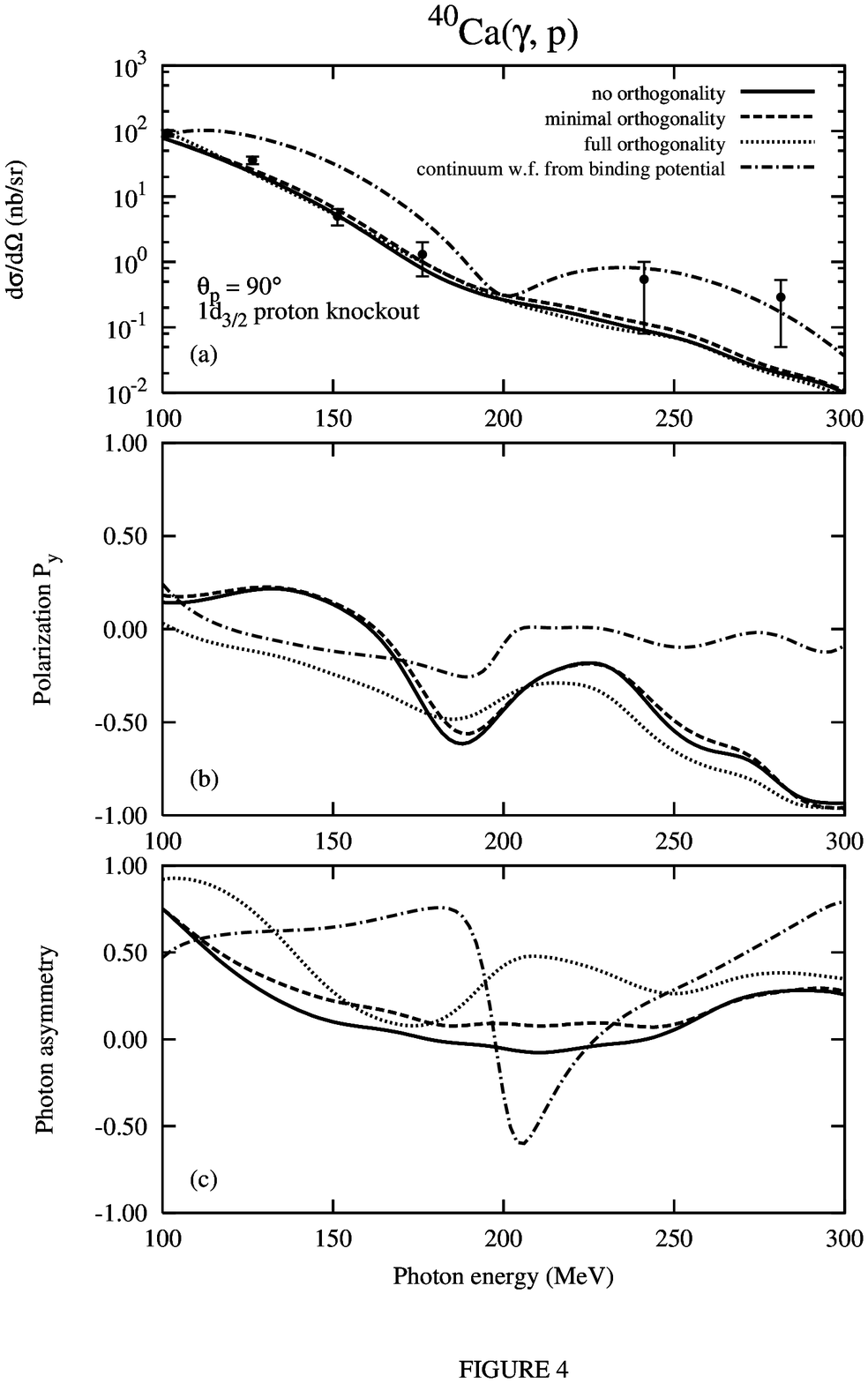}
\end{picture}
\end{figure}

\begin{figure}
\begin{picture}(1100,400)(0,0)
\includegraphics{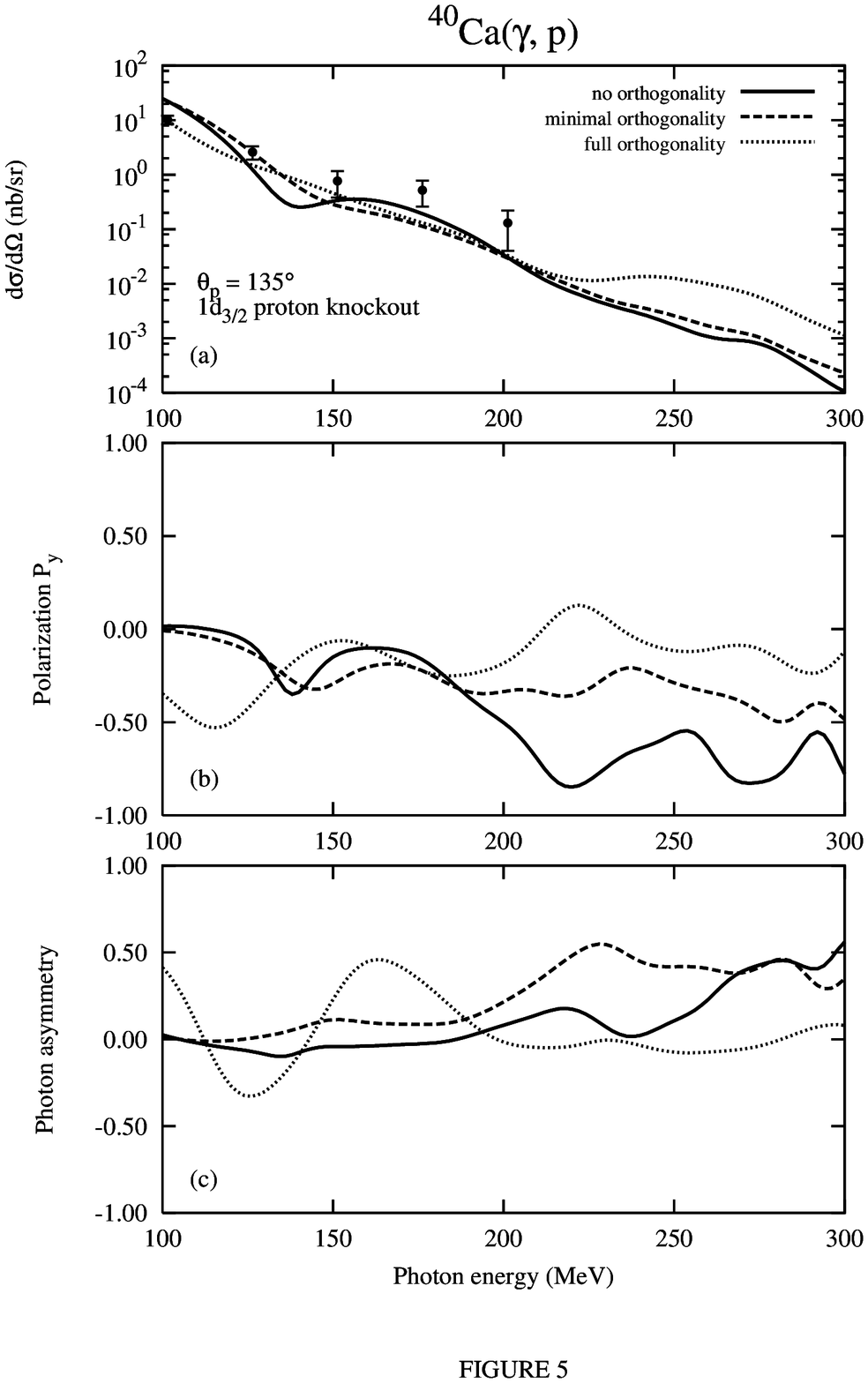}
\end{picture}
\end{figure}

\begin{figure}
\begin{picture}(1100,400)(0,0)
\includegraphics{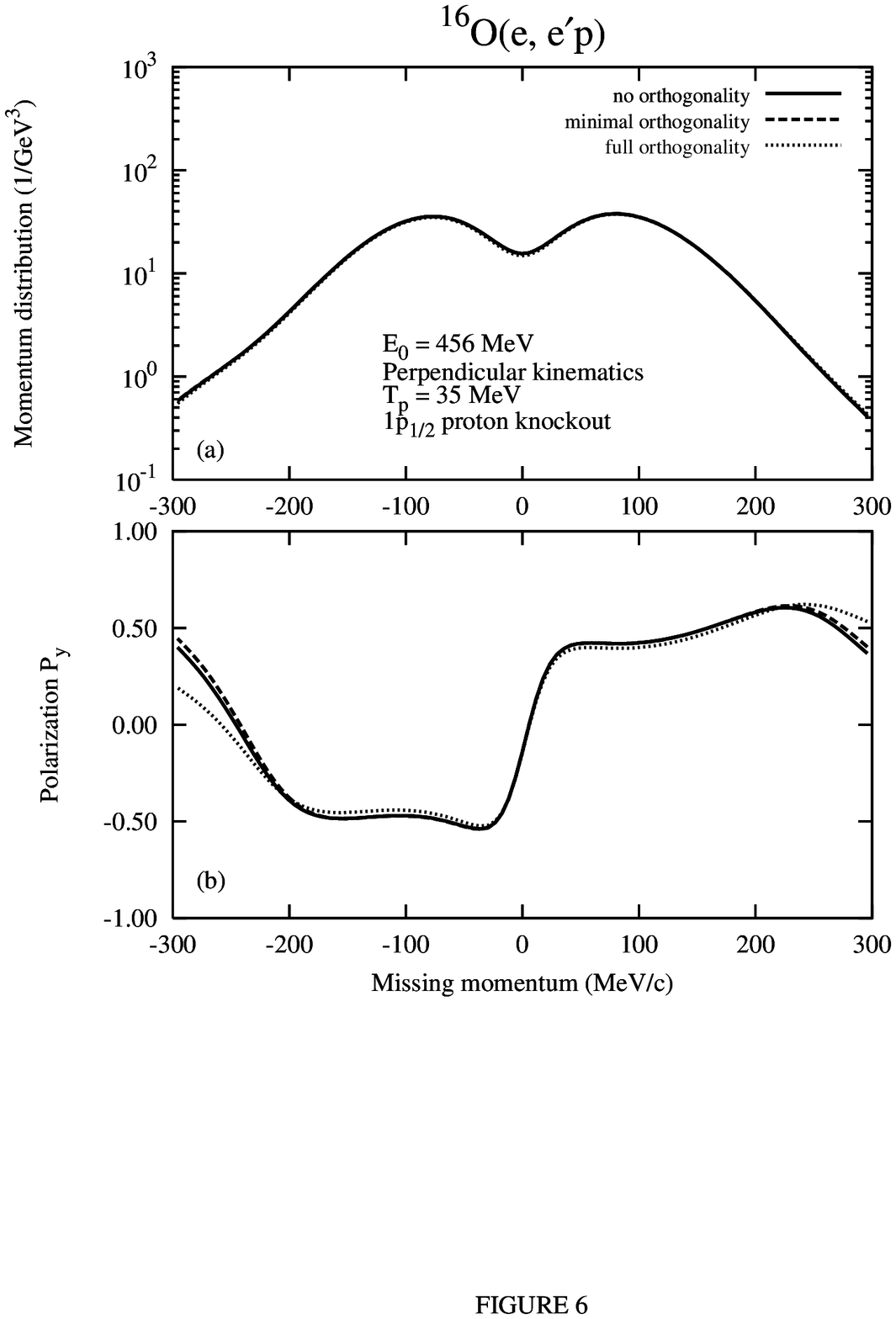}
\end{picture}
\end{figure}

\begin{figure}
\begin{picture}(1100,400)(0,0)
\includegraphics{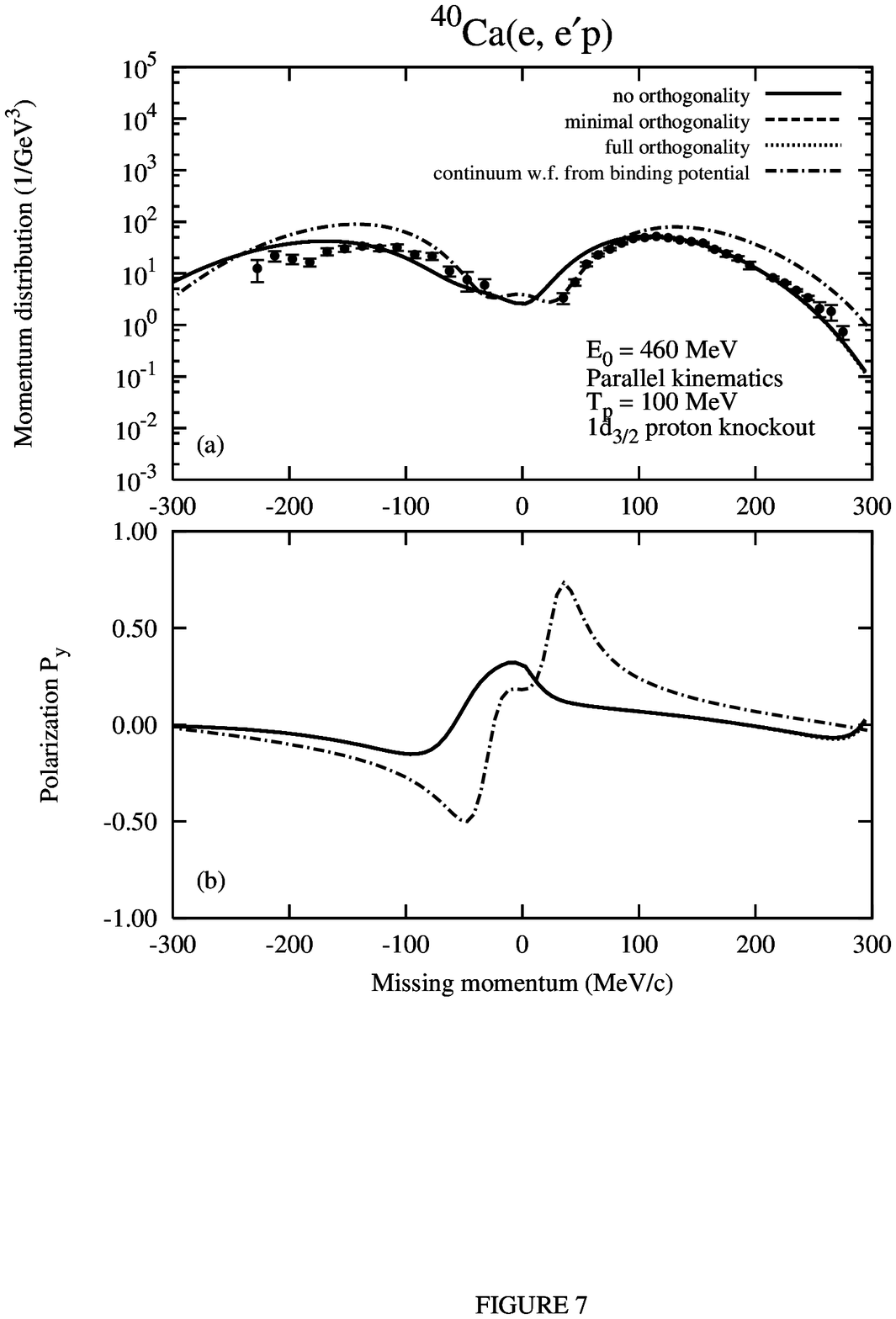}
\end{picture}
\end{figure}

\end{document}